\documentstyle[aps,preprint,epsf,amssymb,graphicx]{revtex}
\tightenlines
\draft
\begin{document}
\title{Inflaton field governed universe
from NKK theory of gravity: stochastic approach.}
\author{$^{1,3}$ Mariano Anabitarte\footnote{
E-mail address: anabitar@mdp.edu.ar}
$^2$Jos\'e Edgar Madriz Aguilar\footnote{
E-mail address: edgar@itzel.ifm.umich.mx}
and $^{1,3}$Mauricio Bellini\footnote{
E-mail address: mbellini@mdp.edu.ar}}
\address{$^1$Departamento de F\'{\i}sica,
Facultad de Ciencias Exactas y Naturales,
Universidad Nacional de Mar del Plata,
Funes 3350, (7600) Mar del Plata, Argentina.\\
$^2$Instituto de F\'{\i}sica y Matem\'aticas,
AP: 2-82, (58040) Universidad Michoacana de San Nicol\'as de Hidalgo,
Morelia, Michoac\'an, M\'exico.\\
$^3$ Consejo Nacional de Ciencia y Tecnolog\'{\i}a (CONICET).}

\vskip .5cm
\maketitle
\begin{abstract}
We study a nonperturbative
single field (inflaton) governed cosmological model from
a 5D Noncompact Kaluza-Klein (NKK) theory of gravity.
The inflaton field fluctuations are estimated for
different epochs of the evolution of the universe.
We conclude that the inflaton field has been sliding down
its (quadratic) potential hill along all the evolution
of the universe and a mass of the order of the Hubble parameter.
In the model here developed the only free parameter is the Hubble
parameter, which could be reconstructed in future from
Super Nova Acceleration Probe (SNAP) data.
\end{abstract}
\vskip .2cm
\noindent
Pacs numbers: 04.20.Jb, 11.10.kk, 98.80.Cq \\
\vskip 1cm
\section{Introduction}

The possibility that space-time had more than four dimensions has
widely been studied regarding its cosmological aspects since long ago\cite{1}.
Investigations
have focused on attempts to explain why the universe presently appears to
have only four space-time dimensions if it is, in fact, a dynamically
evolving $(4+k)$-dimensional manifold ($k$ being
the number of extra dimensions). It has been shown that solutions to the
$(4+k)$-dimensional Einstein equations exist, for which 4D
space-time expands while the extra dimensions contract or remain
constant. It has been also suggested that experimental detection of the time
variation of the fundamental constants could provide strong evidence
for the existence of extra dimensions\cite{2}.
In the past years, there has been a marked resurgence of interest
in models with non-compact or large extra-dimensions. Three examples
of such scenarios are the most known - namely the braneworld models
of Randall and Sundrum (RS)\cite{3}
and Arkani-Hamed, Dimopoulos and Dvali (ADD)\cite{4,4-1},
as well as the older Space-Time-Matter theory (STM)\cite{5}.
The RS model is motivated from certain ideas in string theory,
which suggest that the particles and fields of the standard model are
naturally confined to a lower-dimensional hypersurface living in
a non-compact, higher-dimensional bulk manifold. The driving goal
behind the ADD picture is to explain the discrepancy
in scale between the observed strength of the gravitational
interaction and the other fundamental forces.
This is accomplished by noting that
in generic higher-dimensional models with compact extra dimensions, the
bulk Newton's constant is related to the effective 4D
constant by factors depending on the size and
number of the extra dimensions.
Finally, STM or induced matter theory proposes that our universe is
an embedded 4D-surface in a vacuum
5D-manifold. In this picture, what we perceive to be the source
in the 4D Einstein field equations is really just an
artifact of the
embedding; or in other words, conventional matter is induced from
higher-dimensional geometry.

This paper is devoted to the study of a nonperturbative
single field (inflaton) governed cosmological model from
a 5D NKK theory of gravity.
In a cosmological context, the energy of scalar fields has been argued
to contribute to the expansion of the
universe\cite{vh}, and has been proposed to
explain inflation as well as the presently accelerated
expansion.
We have in mind an universe which initially suffers an inflationary
expansion that after inflation has a change of phase towards a
decelerated expansion (radiation and matter dominated expansions), and
thereinafter evolves towards the present day observed accelerated
(quintessential) expansion.
We consider that the universe is in apparent vacuum on
the 5D globally flat ($R^A_{BCD}=0$) metric.
The 5D apparent vacuum is considered
as a purely kinetic Lagrangian for a scalar field
minimally coupled to gravity on a 5D globally flat metric.

\section{Reviewed and extended formalism}

\subsection{The inflaton field in a 5D vacuum state}

We consider the canonical 5D metric\cite{EPJ04}
\begin{equation}\label{6}
dS^2 = \theta\left(\psi^2 dN^2 - \psi^2 e^{2N} dr^2 - d\psi^2\right),
\end{equation}
where $dr^2=dx^2+dy^2+dz^2$. 
Here, the coordinates ($N$,$\vec r$)
are dimensionless, the fifth coordinate
$\psi $ has spatial units and $\theta$ is a dimensionless parameter
that can take the values $\theta = \pm 1$. 
The metric (\ref{6}) describes a
flat 5D manifold in apparent vacuum ($G_{AB}=0$).
Notice we are considering a diagonal metric because we are dealing only with
gravitational effects, which are the important ones in the global evolution
for the universe. Furthermore, the metric (\ref{6}) is considered
as 3D spatially isotropic and flat: ${r^2 \over 3} = x^2 =y^2 = z^2$
and globally flat ($R^A_{BCD}=0$).

To describe the 5D vacuum universe, we consider an action
\begin{equation} 
I=-\int d^{4}xd\psi\,\sqrt{\left|\frac{^{(5)}
g}{^{(5)}g_0}\right|} \ \left[\frac{^{(5)}R}{16\pi G}+ ^{(5)}{\cal 
L}(\varphi,\varphi_{,A})\right],
\label{action}
\end{equation}
for a scalar field $\varphi$, which is minimally coupled to gravity.
For the metric (\ref{6}), $|^{(5)}g|=\psi^8 e^{6N}$
is the absolute value for the determinant of 
$g_{AB}$ and
$|^{(5)}g_0|=\psi^8_0 e^{6N_0}$ is a constant of dimensionalization 
determined by $|^{(5)}g|$ evaluated at $\psi=\psi_0$ and $N=N_0$.
Furthermore,
$^{(5)}R$ is the 5D Ricci scalar, $G=M^{-2}_p$ is the gravitational constant
and $M_p=1.2 \  10^{19} \  {\rm GeV}$ is the Planckian mass.
In this work we shall consider $N_0=0$, so that
$\left|^{(5)}g_0\right|=\psi^8_0$. Here, the index $`` 0 "$ denotes the value
at the end of inflation (i.e.,
when $\ddot b =0$).
Since we are aimed to describe a manifold in apparent vacuum
the Lagrangian density ${\cal L}$ in (\ref{action}) should be only
kinetic in origin
\begin{equation}\label{1'}
^{(5)}{\cal L}(\varphi,\varphi_{,A}) =
\frac{1}{2} g^{AB} \varphi_{,A} \varphi_{,B},
\end{equation}
where the diagonal tensor metric $g^{AB}$ is given by
the line element (\ref{6}).

Since
${\partial N \over \partial\psi}$ and ${\partial\psi \over \partial N}$
are zero (the coordinates are independent), 
the equation of motion for the scalar quantum field $\varphi$ is
\begin{equation}
\stackrel{\star \star}{\varphi}+3\stackrel{\star}{\varphi}-e^{-2 
N}\nabla_{r}^{2}\varphi -\left[4\psi 
\frac{\partial\varphi}{\partial\psi}+\psi^{2} 
\frac{\partial^{2}\varphi}{\partial\psi^{2}}\right]=0,
\label{ec2}
\end{equation}
where the overstar denotes the derivative with respect to $N$ and 
$\varphi \equiv \varphi(N,\vec{r},\psi)$.
The commutator between
$\varphi$ and $\Pi^N = {\partial {\cal L} \over \partial
\varphi_{,N}} = g^{NN} \varphi_{,N}$ is given by
\begin{equation}
\left[\varphi(N,\vec r,\psi), \Pi^N(N,\vec{r'},\psi')\right] =
ig^{NN} \left|\frac{^{(5)} g_0}{^{(5)} g}\right| \  \delta^{(3)}(\vec r-
\vec{r'}) \delta(\psi - \psi'),
\end{equation}
where $\left|{^{(5)} g_0 \over ^{(5)} g}\right|$ is the inverse
of the normalized volume of the manifold (\ref{6}) and $g^{NN} = \psi^{-2}$.
By means of the transformation $\varphi =\chi 
e^{-3N/2}\left(\psi_{0}\over \psi \right)^{2}$ we obtain the 
5D generalized Klein-Gordon like equation for 
$\chi (N,\vec{r},\psi)$ and the commutator between $\chi$ and
$\stackrel{\star}{\chi}$:
\begin{eqnarray}
&& \stackrel{\star \star}{\chi}-\left[e^{-2 N}\nabla_{r}^{2} 
+\left(\psi^{2} \frac{\partial^2}{\partial\psi^2} +\frac{1}{4}\right) 
\right]\chi =0,  \label{ec3} \\
&& 
\left[\chi(N,\vec r,\psi), \stackrel{\star}{\chi}
(N,\vec{r'},\psi')\right] =
i  \delta^{(3)}(\vec r- \vec{r'}) \delta(\psi - \psi'). \label{commm}
\end{eqnarray}

The redefined field $\chi$ can be written in terms of a Fourier expansion
in terms of the modes $\chi_{k_r k_{\psi}}(N, \vec r, \psi)
=e^{i(\vec{k_r} \cdot \vec{r}
+{k_\psi}\cdot {\psi})}\xi_{k_{r}k_{\psi}}(N,\psi)$
\begin{eqnarray} \label{ec4}
\chi (N,\vec{r},\psi)&=& \frac{1}{(2\pi)^{3/2}}\int d^{3} k_{r} \int d 
k_{\psi} \left[a_{k_{r}k_{\psi}} e^{i(\vec{k_r} \cdot \vec{r} 
+{k_\psi}\cdot {\psi})}\xi_{k_{r}k_{\psi}}(N,\psi) \right. \nonumber
\\
&+& \left. a_{k_{r}k_{\psi}}^{\dagger} e^{-i(\vec{k_r} \cdot \vec{r} 
+{k_\psi} \cdot {\psi})}\xi_{k_{r}k_{\psi}}^{*}(N,\psi)\right],
\end{eqnarray}
where the asterisk denotes the complex conjugate and 
$(a_{k_{r}k_{\psi}},a_{k_{r}k_{\psi}}^{\dagger})$ are respectively
the annihilation
and creation operators which satisfy the following commutation
expressions
\begin{eqnarray}\label{ec5}
\left[a_{k_{r}k_{\psi}},a_{k'_{r}k'_{\psi}}^{\dagger}\right]&=&\delta^{(3)
}\left(\vec{k_r}-\vec{k'_r}\right) 
\delta\left(\vec{k_\psi}-\vec{k'_\psi}\right),\\
\label{ec6}
\left[a_{k_{r}k_{\psi}}^{\dagger},a_{k'_{r}k'_{\psi}}^{\dagger}\right]
&=&\left[a_{k_{r}k_{\psi}},a_{k'_{r}k'_{\psi}}\right]=0.
\end{eqnarray}
The expression (\ref{commm}) complies if the modes are normalized
by the following condition:
\begin{equation} \label{recon}
\xi_{k_{r}k_{\psi}}\left(\stackrel{\star}{\xi}_{k_{r} 
k_{\psi}}\right)^{*} - \left(\xi_{k_{r} k_{\psi}}\right)^{*} 
\stackrel{\star}{\xi}_{k_{r}k_{ \psi }} = i.
\end{equation}
This equation provides the normalization for the complete set of
solutions on all the ($k_r,k_{\psi}$) spectrum.
On the other hand, the dynamics for the
modes $\xi_{k_{r}k_{\psi}}(N,\psi)$ is given by
the equation
\begin{equation} \label{ec8}
\stackrel{\star \star}{\xi}_{k_{r}k_{\psi}} +k_{r}^{2} e^{-2 N} 
\xi_{k_{r}k_{\psi}} +\psi^{2}\left(k_{\psi}^{2} 
-2ik_{\psi}\frac{\partial}{\partial \psi}-\frac{\partial^{2}}{\partial 
\psi^{2}} -\frac{1}{4\psi^{2}}\right)\xi_{k_{r}k_{\psi}}=0.
\end{equation}

The solution of (\ref{ec8}) with the condition (\ref{recon})
can be written as\cite{gi27}
\begin{equation} \label{ec17}
\xi_{k_{r}k_{\psi}}(N,\psi)=\frac{i\sqrt{\pi}}{2}e^{-i\vec{k}_{\psi} 
\cdot \vec{\psi}}{\mathcal H}_{1/2}^{(2)}[k_{r}e^{-N}]=
e^{-i\vec{k}_{\psi}.\vec\psi} \  \bar\xi_{k_r}(N),
\end{equation}
where ${\mathcal H}_{1/2}^{(1,2)}[x(N)]={\mathcal J}_{1/2}[x(N)]\pm 
i{\mathcal Y}_{1/2}[x(N)]$ are the Hankel functions, ${\mathcal 
J}_{1/2}[x(N)]$ and ${\mathcal Y}_{1/2}[x(N)]$ are the first and second kind 
Bessel functions with $x(N)=k_{r} e^{-N}$. 
Furthermore the function $\bar{\xi}_{k_r}(N)$
is given by
\begin{equation}
\bar{\xi}_{k_r}(N)
=\frac{i \sqrt{\pi}}{2} {\cal H}^{(2)}_{1/2}
\left[ k_r e^{-N}\right],
\end{equation}
such that the normalization condition for $\bar\xi_{k_r}(N)$ becomes
\begin{equation}
\bar\xi_{k_{r}}\left(\stackrel{\star}{\bar\xi}_{k_{r} 
}\right)^{*} - \left(\bar\xi_{k_{r}}\right)^{*} 
\stackrel{\star}{\bar\xi}_{k_{r}} = i.
\end{equation}

Finally, the field $\chi$ in eq. (\ref{ec4}) can be
rewritten as
\begin{equation}\label{chichi}
\chi(N,\vec r,\psi) =\chi(N,\vec r)
=\frac{1}{(2\pi)^{3/2}} {\Large\int} d^3k_r
{\Large\int} dk_{\psi} \left[ a_{k_r k_{\psi}} e^{i \vec{k_r}.\vec r}
\bar\xi_{k_r}(N) + a^{\dagger}_{k_r k_{\psi}} e^{-i \vec{k_r}.\vec r}
\bar\xi^*_{k_r}(N)\right],
\end{equation}
and the  field $\varphi$ is given by
\begin{equation}
\varphi(N,\vec r,\psi) = e^{-\frac{3N}{2}}
\left(\frac{\psi_0}{\psi}\right)^2
\chi(N,\vec r),
\end{equation}
with $\chi(N,\vec r)$ given by eq. (\ref{chichi}).
Note that exponentials $e^{\pm i \vec{k}_{\psi}.\vec{\psi}}$ disappear
in $\chi(N,\vec r)$ and there is not dependence on the fifth coordinate
$\psi$ in this field. This is a very important fact that says us
that the field $\varphi(N, \vec r, \psi)$ propagates only on the
3D spatially isotropic space $r(x,y,z)$, but not on the additional
space-like coordinate $\psi$. Hence, gravity should not be localized
on the fifth dimension and therefore the usual Newton law should not
hold on the 5D manifold.

\subsection{Coarse-granning of $\varphi$ in a 5D vacuum state}

To study the evolution of the field $\varphi$ on
large 3D spatial scales, we can introduce the field $\chi_L$
\begin{equation}
\chi_L(N,\vec r) =
\frac{1}{(2\pi)^{3/2}} {\Large\int} d^3k_r
{\Large\int} dk_{\psi} \  \Theta(\epsilon k_0(N) -k_r)
\left[ a_{k_r k_{\psi}} e^{i \vec{k_r}.\vec r}
\bar\xi_{k_r}(N) + c.c.\right],
\end{equation}
where $\Theta$ denotes the Heaviside function.
Furthermore, $c.c.$ denotes the complex conjugate of the first term inside
the brackets and $k_0=1/2 e^N $
is the $N$-dependent wavenumber (related to the 3D spatially isotropic,
homogeneous and flat space $r^2=x^2+y^2+z^2$), which
separates the long ($k^2_r \ll k^2_0$) and short ($k^2_r \gg k^2_0$)
wavelength sectors. Modes with $k_r/k_0 <\epsilon$ are referred to as
outside the horizon.

If the short wavelenght modes are described with the field $\chi_S$
\begin{equation}
\chi_S(N,\vec r) = \frac{1}{(2\pi)^{3/2}} {\Large\int} d^3k_r
{\Large\int} dk_{\psi} \Theta(k_r-\epsilon k_0(N))
\left[ a_{k_r k_{\psi}} e^{i \vec{k_r}.\vec r}
\bar\xi_{k_r}(N) + c.c.\right],
\end{equation}
such that $\chi=\chi_L + \chi_S$, hence
the equation of motion for $\chi_L$
will be approximately
\begin{equation}\label{sto1}
\stackrel{\star\star}{\chi}_L - \left(\frac{k_0(N) b_0}{b}\right)^2 \chi_L =
\epsilon \left[\stackrel{\star\star}{k_0} \eta(N,\vec r) +
\stackrel{\star}{k_0} \kappa(N,\vec r) + 2 \stackrel{\star}{k_0}
\gamma(N,\vec r)\right],
\end{equation}
where the stochastic operators $\eta$, $\kappa$ and $\gamma$ are given
respectively by
\begin{eqnarray}
\eta(N,\vec r) &=&
\frac{1}{(2\pi)^{3/2}} {\Large\int} d^3k_r
{\Large\int} dk_{\psi} \  \delta(\epsilon k_0(N) -k_r)
\left[ a_{k_r k_{\psi}} e^{i \vec{k_r}.\vec r}
\bar\xi_{k_r}(N) + c.c.\right], \\
\kappa(N,\vec r) & =&
\frac{1}{(2\pi)^{3/2}} {\Large\int} d^3k_r
{\Large\int} dk_{\psi} \  \stackrel{\star}{\delta}(\epsilon k_0(N) -k_r)
\left[ a_{k_r k_{\psi}} e^{i \vec{k_r}.\vec r}
\bar\xi_{k_r}(N) + c.c.\right],\\
\gamma(N,\vec r) & = &
\frac{1}{(2\pi)^{3/2}} {\Large\int} d^3k_r
{\Large\int} dk_{\psi} \  \delta(\epsilon k_0(N) -k_r)
\left[ a_{k_r k_{\psi}} e^{i \vec{k_r}.\vec r}
\stackrel{\star}{\bar\xi}_{k_r}(N) + c.c.\right].
\end{eqnarray}
The equation (\ref{sto1}) can be rewritten as
\begin{equation}\label{sto2}
\stackrel{\star\star}{\chi}_L - \frac{1}{4} \chi_L =
\epsilon \left[\frac{d}{dN}\left(\stackrel{\star}{k_0}
\eta(N,\vec r)\right) +
\stackrel{\star}{k_0} \gamma(N,\vec r)\right].
\end{equation}
This is a
Kramers-like stochastic equation that can be written as two
first order stochastic (Langevin) ones by introducing
the auxiliar field
$u= \stackrel{\star}{\chi_L} - \epsilon \stackrel{\star}{k_0}\eta$
\begin{eqnarray}
\stackrel{\star}{u} &=& \frac{1}{4} \chi_L +
\epsilon \stackrel{\star}{k_0} \gamma, \label{uno} \\
\stackrel{\star}{\chi}_L &=&
u + \epsilon \stackrel{\star}{k_0} \eta.\label{dos}
\end{eqnarray}
The role of the
noise $\gamma$ can be minimized in the system (\ref{uno})-(\ref{dos}) 
if $\left(\stackrel{\star}{k_0}\right)^2
\left<\gamma^2\right> \ll \left(\stackrel{\star\star}{k_0}\right)^2
\left<\eta^2\right>$, which holds
when
\begin{equation}\label{condition}
\frac{\stackrel{\star}{\bar\xi_{k_r}}
\left(\stackrel{\star}{\bar\xi}_{k_r}\right)^*}{
\bar\xi_{k_r} \left(\bar\xi_{k_r}\right)^* }\ll 1.
\end{equation}
In that case the
system (\ref{uno}), (\ref{dos}) can be approximated to
\begin{eqnarray}
\stackrel{\star}{u} &=& \alpha \chi_L , \label{uno1} \\
\stackrel{\star}{\chi}_L &=& u + \epsilon \stackrel{\star}{k_0}
\eta.\label{dos2}
\end{eqnarray}
This system represents two Langevin equations with a noise
$\eta$ which is gaussian and white in nature
\begin{eqnarray}
&& \left< \eta\right> = 0, \\
&& \left< \eta^2\right> = \frac{\epsilon \left(k_0\right)^2}{2\pi^2
\stackrel{\star}{k_0}} {\Large\int} dk_{\psi} \  \bar\xi_{\epsilon k_0}
\bar\xi^*_{\epsilon k_0} \  \delta(N-N').
\end{eqnarray}
The equation that describes the dynamics of the transition probability
$P\left[\chi^{(0)}_L, u^{(0)}|\chi_L, u\right]$
from a configuration ($\chi^{(0)}_L, u^{(0)}$) to ($\chi_L, u$) is
a Fokker-Planck one
\begin{equation}
\frac{\partial P}{\partial N} = - u \frac{\partial P}{\partial\chi_L}
-\frac{1}{4} \chi_L \frac{\partial P}{\partial u}
+ \frac{1}{2} D_{11} \frac{\partial^2 P}{\partial \chi_L^2},
\end{equation}
where $D_{11}={1 \over 2} \left(\epsilon \stackrel{\star}{k_0}\right)^2
\left[\int dN \left<\eta^2\right>\right]$
is the diffusion coefficient related to the variable $\chi_L$.
Explicitely
\begin{equation}
D_{11} = \frac{\epsilon^3 \left(k_0\right)^2}{4\pi^2}
\stackrel{\star}{k_0} {\Large\int} dk_{\psi} \  \bar\xi_{\epsilon k_0}
\bar\xi^*_{\epsilon k_0},
\end{equation}
which is divergent.\\

\subsection{The metric}

Now we consider the 5D canonical metric (\ref{6}).
In the 3D comoving frame $U^r=0$,
the geodesic dynamics ${dU^C \over dS}=-\Gamma^C_{AB} U^A U^B$
with $g_{AB} U^A U^B=1$, give us the velocities $U^A$ (latin letters
take values $0,1,2,3,4$)
\begin{equation}\label{u}
U^N={u(N) \over \psi\sqrt{u^2(N)-1}}, \qquad U^{r}=0, \qquad
U^{\psi} = - {1 \over \sqrt{u^2(N)-1}},
\end{equation}
which are satisfied for $S(N)=\pm |N|$.
In this work we shall consider the case $S(N) = |N|$.
Note that the solution (\ref{u}) is one of the possible representations
of the general solution
\begin{eqnarray}
&& U^N = \frac{{\rm cosh}[S(N)]}{\psi}, \\
&& U^r =0, \\
&& U^{\psi} = - {\rm sinh}[S(N)].
\end{eqnarray}
In the representation (\ref{u}) we obtain
${d\psi \over dN}\equiv {U^{\psi}\over U^N}
=\psi/u(N)$, where
$u(N)$ is an arbitrary function such that ${\rm tanh}[S(N)] = -1/u(N)$.
Thus, the fifth coordinate evolves as
\begin{equation}\label{psi}
\psi(N) = \psi_0 e^{\int dN/u(N)}.
\end{equation}
Here, $\psi_0$ is a constant of integration that has spatial units.
From the mathematical point of view, we are taking a foliation
of the 5D metric (\ref{6}) with $r$ constant.
Hence, to describe
the metric in physical coordinates we must make the
following transformations\cite{tt}:
\begin{equation}\label{tran}
t = \int \psi(N) dN, \qquad R=r\psi, \qquad L= \psi(N) \  e^{-\int dN/u(N)},
\end{equation}
such that for $\psi(t)=1/h(t)$,
we obtain the 5D metric
\begin{equation}\label{m1}
dS^2 = \theta\left(dt^2 - e^{2\int h(t) dt} dR^2 - dL^2\right),
\end{equation}
where $L=\psi_0$ is a constant and $h(t)=\dot b/b$ is the effective
Hubble parameter and $b$ is the effective scale factor of the
universe.
The metric (\ref{m1}) describes a 5D generalized FRW
metric, which is 3D spatially flat [i.e., it is flat in terms of
$\vec R = (X, Y, Z)$], isotropic and homogeneous.
In the representation $(\vec R,t,L)$, the 
velocities $ \hat U^A ={\partial \hat x^A \over \partial x^B} U^B$,
are
\begin{equation} \label{10}
\hat U^t=\frac{2u(t)}{\sqrt{u^2(t)-1}}, \qquad
\hat U^R=-\frac{2R h}{\sqrt{u^2(t)-1}}, \qquad \hat U^L=0,
\end{equation}
where
the old velocities $U^B$ are $U^N$, $U^r=0$ and $U^{\psi}$
and the velocities $\hat U^B$ are constrained by the condition
\begin{equation}\label{con}
\hat g_{AB} \hat U^A \hat U^B =1.
\end{equation}
Note that the metric (\ref{m1}) is not globally flat and the line
element (\ref{m1}) do not
describes a 5D vacuum state.
Furthermore, we are considering an observer which is in the frame
described by the velocities (\ref{10}). To avoid confusion, notice
that the transformation (\ref{tran}) only describes a map from the
particular frame (\ref{u}) [of the metric ({\ref{6})], to the particular
frame (\ref{10}) [of the metric (\ref{m1})]. However, the transformation
(\ref{tran}) is not a general map from the metric (\ref{6}) to (\ref{m1}).
Hence, if an observer is in the frame (\ref{10}), he only see
the effective 4D FRW metric
\begin{equation}\label{frw}
dS^2 \rightarrow ds^2 = \theta
\left(dt^2 - e^{2\int h(t) dt} dR^2\right),
\end{equation}
which has an effective
4D scalar curvature $^{(4)}{\cal R} = 6(\dot h + 2 h^2)$. The
metric (\ref{frw}) has a metric tensor with components $g_{\mu\nu}$
($\mu,\nu$ take the values $0,1,2,3$).
The absolute value of the
determinant for this tensor is $\left|^{(4)}g\right|
=(b/b_0)^6$.
Note that $\phi_0$ not necessarily takes the usual 4D Planckian length.
Furthermore, the 4D energy
density $\rho$ and the pressure ${\rm p}$ are
\begin{eqnarray}
&& 8 \pi G \rho = 3 h^2,\\
&& 8\pi G {\rm p} = -(3h^2 + 2 \dot h),
\end{eqnarray}
where $\dot h <0$ during all the evolution of the universe.

The function $u$ can be written as a function of the cosmic time
$u(t) = -{h^2 \over \dot h}$,
where the overdot represents the derivative with respect to $t$.
The solution $N={\rm arctanh}[1/u(t)]$ corresponds to a
time dependent power-law expanding universe
$h(t)=p(t) t^{-1}$, such that the effective scale factor goes as
$b \sim e^{\int p(t)/t dt}$.
When $u^2(t) >1$, the velocities $\hat U^t$ and $\hat U^R$ are real, so that
the condition (\ref{con}) implies that $\theta =1$. [Note that the
function $u(t)$ can be related to the deceleration parameter
$q(t) = -\ddot b b/\dot b^2$: $u(t) = 1/[1+q(t)]$.]
In such a case the expansion of the universe is accelerated ($\ddot b >0$).
However, when $u^2 <1$ the velocities $U^t$ and $U^R$ are imaginary
and the condition (\ref{con}) holds for $\theta = -1$.
In this case the expansion of the universe is decelerated because
$\ddot b <0$. So, the parameter $\theta$ is introduced in the metric
(\ref{m1}) to preserve the hyperbolic condition (\ref{con}).
The new coordinate $R$ gives us
the physical distance between galaxies separated
by cosmological distances: $R(t)=r(t)/h(t)$,
where $r(t)$ is given by $3 u^2(t) = 4 r^2(t) \left(b/b_0\right)^2 -1$:
\begin{equation}\label{r}
r^2(t) = \left[\frac{3}{4} \frac{h^4}{\dot h^2} + \frac{1}{4}\right]
e^{-2\int h(t) dt},
\end{equation}
for a given evolution of the universe described with $b(t)$.
For $r >1$ ($r <1$), the 3D spatial distance $R(t)$ is defined
on super (sub) Hubble scales.\\

\subsection{The inflaton field in an effective 4D FRW metric}

The 4D Lagrangian corresponding to the
effective 4D Friedmann-Robertson-Walker (FRW) metric (\ref{frw})
[i.e., in the frame (\ref{10})], is given by
\begin{equation}
^{(4)}{\cal L}(\varphi,\varphi_{,\mu}) =
-\sqrt{\left|\frac{^{(4)}g}{^{(4)}g_0}\right|} \left[
\frac{1}{2} g^{\mu\nu} \varphi_{,\mu}\varphi_{,\nu} + V(\varphi)\right],
\end{equation}
where the effective potential for the 4D FRW metric (\ref{frw}), is
\begin{equation}\label{pot}
V(\varphi) = -\left.\frac{1}{2} g^{\psi\psi} \varphi_{,\psi}\varphi_{,\psi}
\right|_{\psi=h^{-1}} =
\frac{1}{2} \left.
\left(\frac{\partial\varphi}{\partial\psi}\right)^2\right|_{\psi=h^{-1}}.
\end{equation}
In our case this potential in the frame (\ref{10}) and $\psi=1/h$,
takes the form
\begin{equation}\label{pot2}
V(\varphi) = 2 h^2(t) \  \varphi^2(t,\vec R,L).
\end{equation}
Notice this potential has a geometrical origin and takes different
representations in different frames. In our case the observer is in a
frame $\hat U^{L}=0$, because we are taking a foliation $L=\psi_0$
on the 5D metric (\ref{m1}).
Furthermore, the effective 4D equation of motion for $\varphi$ is
\begin{equation}\label{mo}
\ddot{\varphi} + \left(3h-\frac{\dot h}{h}\right)
\dot{\varphi} - e^{-2\int 
h(t)\,dt}\nabla_{R}^{2}\varphi - 
\left. \left[\frac{4}{\psi} \frac{\partial\varphi}{\partial\psi}
+ \frac{\partial^2\varphi}{\partial\psi^2}\right]
\right|_{\psi=h^{-1}} = 0,
\end{equation}
which means that the effective 4D
expression for ${dV(\varphi) \over d\varphi}$
is
\begin{equation}
\left.V'(\varphi)\right|_{\psi=h^{-1}} =
2 h^2(t) \  \varphi(\vec R,t,L) - \frac{\dot h}{h} \dot\varphi(\vec R,t,L).
\end{equation}

In order to simplify the structure of the equation (\ref{mo})
we can make the following transformation:
\begin{equation}
\varphi(\vec R, t) =
e^{-\frac{1}{2}\int \left(3h-\dot h/h\right) dt} \chi (\vec R, t),
\end{equation}
such that we obtain the following 4D Klein-Gordon equation
for $\chi $
\begin{equation}   \label{mod}
\ddot\chi - \left[e^{-2\int h(t)dt} \nabla^2_R +
\frac{h^2}{4} + \frac{3}{4} \left(\frac{\dot h}{h}\right)^2 - \frac{1}{2}
\frac{\ddot h}{h} \right]\chi =0.
\end{equation}
The field $\chi$ can be expanded as a Fourier's representation
in terms of the modes $\chi_{k_R k_L}(\vec R,t) = e^{i \vec{k_R}.\vec{ R}}
\bar\xi_{k_R}(t)$
\begin{equation}
\chi(\vec R, t) = \frac{1}{(2\pi)^{3/2}}
{\Large\int} d^3 k_R {\Large\int} dk_{L} \left[
a_{k_R k_{L}} e^{i \vec{k_R}.\vec{ R}}
\bar\xi_{k_R}(t) + c.c.
\right] \  \delta\left( k_{L} - k_{\psi_0}\right),
\end{equation}
where the dynamics for
the modes $\bar\xi_{k_R}(t)$ is given by
\begin{equation}\label{mot}
\ddot{\bar{\xi}}_{k_R} + \left[ k^2_R e^{-2\int h(t)dt} 
-\frac{h^2}{4} - \frac{3}{4} \left(\frac{\dot h}{h}\right)^2
+\frac{1}{2} \frac{\ddot h}{h}\right] \bar{\xi}_{k_R } =0.
\end{equation}
It is important to notice that eq. (\ref{mot}) is exactly the equation
for $\bar\xi_{k_r}(N)$ with the variables transformation (\ref{tran}),
on the hypersurface $\psi=h^{-1}$.\\

\subsection{Coarse-granning of $\varphi$
in an effective 4D cosmological metric}

Now we can define the fields $\chi_L(t,\vec R)$
and $\chi_S(t,\vec R)$, which describe respectively the
long and short wavelength sectors of the field $\chi$
on the effective 4D FRW metric (\ref{frw})
\begin{eqnarray}
\chi_L(t, \vec R) &=& \frac{1}{(2\pi)^{3/2}} {\Large\int} d^3k_R
{\Large\int}dk_L \  \Theta(\epsilon F(t) - k_R) \left[
a_{k_R k_{\psi}} e^{i \vec{k_R}.\vec{R}} \bar\xi_{k_R}(t) + c.c.\right]
\delta(k_{L} - k_{\psi_0}), \\
\chi_S(t, \vec R) &=& \frac{1}{(2\pi)^{3/2}} {\Large\int} d^3k_R
{\Large\int}dk_L \  \Theta(k_R- \epsilon F(t)) \left[
a_{k_R k_{\psi}} e^{i \vec{k_R}.\vec{R}} \bar\xi_{k_R}(t) + c.c.\right]
\delta(k_{L} - k_{\psi_0}),
\end{eqnarray}
where $F(t) = h(t) e^{\int h dt}$ is the inverse of the Hubble
horizon in an expanding universe.
The field
that describes the dynamics of $\chi$ on the infrared sector
($k_R < \epsilon F$) is $\chi_L$. During the inflationary expansion
the dimensionless parameter take values of the order of $10^{-3} - 10^{-4}$.
However, the present day value for $\epsilon$ should be of the order of
$10^2$.
The dynamics of $\chi_L$ obeys
the Kramers-like stochastic equation
\begin{equation}
\ddot\chi_L - \frac{k^2_0 b^2_0}{b^2} \chi_L =
\epsilon \left[ \frac{d}{dt}\left(\dot F \eta(t,\vec R) \right)+
\dot{F} \gamma(t,\vec R)\right],
\end{equation}
where $k_0(t) =e^{\int h dt} \left[{h^2 \over 4}
+ {3\over 4} \left(\dot h/h\right)^2
-{1\over 2} \ddot h/h\right]^{1/2}$
and the stochastic operators $\eta$, $\kappa$ and $\gamma$ are
\begin{eqnarray}
\eta &=& \frac{1}{(2\pi)^{3/2}} {\Large\int} d^3 k_R \  \delta(\epsilon F-k_R)
\left[a_{k_R k_{\psi_0}} e^{i \vec{k_R}.\vec{R}} \bar\xi_{k_R}(t)
+ c.c.\right],\\
\gamma &=& \frac{1}{(2\pi)^{3/2}}
{\Large\int} d^3 k_R \  \delta(\epsilon F-k_R)
\left[a_{k_R k_{\psi_0}} e^{i \vec{k_R}.\vec{R}} \dot{\bar\xi}_{k_R}(t)
+ c.c.\right].
\end{eqnarray}
This second order stochastic equation can be rewritten as two
Langevin stochastic equations
\begin{eqnarray}
&& \dot u = \frac{k^2_0 b^2_0}{b^2} \chi_L
+ \epsilon \dot F \gamma, \label{58} \\
&& \dot\chi_L = u + \epsilon \dot F \eta,
\end{eqnarray}
where $u=\dot\chi_L - \epsilon \dot{F} \gamma$.
The condition to neglect the noise $\gamma$ with respect to
$\eta$, now holds
\begin{equation}\label{ju}
\frac{\dot{\bar\xi}_{k_R} \dot{\bar\xi}^*_{k_R}}{
\bar\xi_{k_R} \bar\xi^*_{k_R}} \ll \frac{\left(\ddot{F}\right)^2}{
\left( \dot{F}\right)^2},
\end{equation}
Notice this result is exactly the same in
eq. (\ref{condition}), with the transformation (\ref{tran}).
The Fokker-Planck equation for the transition probability
$\bar P(\chi^{(0)}_L,u^{(0)}|\chi_L,u)$ is
\begin{equation}
\frac{\partial \bar P}{\partial t} = - u \frac{\partial \bar P}{\partial \chi_L}
-\frac{k^2_0 b^2_0}{b^2} \chi_L \frac{\partial \bar P}{\partial
u} + D_{11}(t) \frac{\partial^2 \bar P}{\partial\chi^2_L},
\end{equation}
where $\bar D_{11}(t) = {\epsilon^3\dot F k^2_0\over
4\pi^2 }\left|\bar\xi_{\epsilon F}\right|^2$.
Hence, the equation of motion for $\left< \chi^2_L\right> =
\int d\chi_L du \chi^2_L \bar P(\chi_L,u)$ is
\begin{equation}\label{mot1}
\frac{d}{dt}\left<\chi^2_L\right>
= \bar D_{11}(t).
\end{equation}
When $\bar D_{11} >0$, $\chi_L$ increases its number of degrees of
freedom. 
In our case the number of degrees of freedom changes depending on
the rate of expansion of the universe. 
In order to return to the original field $\varphi_L =
e^{-{1\over 2} \int\left(3h-{\dot h\over h}\right)} \chi_L$
the equation (\ref{mot1}) can be
rewritten as
\begin{equation}
\frac{d}{dt}\left<\varphi^2_L\right> = -\left(3h -\frac{\dot h}{h}\right)
\left<\varphi^2_L
\right> + \bar D_{11}(t) e^{-\int\left(3h-\frac{\dot h}{h}\right)dt},
\end{equation}
which has the following general solution
\begin{equation}  \label{gens}
\left<\varphi^2_L\right> = e^{-\int^t \left(3h - \frac{\dot h}{h}\right)dt''}
\left[{\Large\int}^t \bar D_{11}(t') dt' + C\right],
\end{equation}
where $C$ is a constant of integration, $\bar\xi_{k_R=\epsilon F}$ is the
solution of eq. (\ref{mot}) with $k_R=\epsilon F$.

\section{Evolution of the universe: a model}

With the aim to study an effective model for the expansion of the universe
we consider a Hubble parameter $h(t) = p(t)/t$, such that the time
dependent power expansion $p(t)$ is given by
\begin{equation}\label{pow}
p(t) = 1.8 \  a t^{-n} - 1.8 \  b t^{-n/2} +
\left(\frac{b^2}{4a} + \frac{2}{3}\right) + c t,
\end{equation}
where $a = 1/6 \  10^{30 n} \  G^{n/2}$, $b = 1/3 \  10^{15n} \  G^{n/4}$,
$c = 10^{-61} \  G^{-1/2}$ and $n=0.352$.
This model represents an early inflationary expansion followed by a decelerated
(radiation dominated followed by a matter dominated expansion) that finally
suffers a present day quintessential accelerated expansion.
Other cosmological models without inflationary expansion has
been considered recently in the literature\cite{Liu}.
The power of expansion (\ref{pow}) is shown in the figure (1),
being $x(t) = log_{10}[t/t_0]$. Note that for $x<10$, $p(t)$ suffers
an inflationary expansion. For $60.135 > x >10$, the universe is decelerated,
being radiation dominated for $x\simeq 30$ and matter dominated
for $x\simeq 55$. For $x > 60.135$ the universe suffers a quintessential
expansion, being its actual age $x \simeq 60.653$ (i.e., approximately
$t \simeq 1.5 \  10^{10}$ years old), which has been experimentally
observed from the supernova (SNe) Type data\cite{a,b,c,d}.
The model here studied
gives us a present day deceleration parameter $q=-\ddot b b/\dot b^2$:
$q(x= 60.653) \simeq -0.492$,
which is in good agreement with experimentation\cite{PDG}.
The figure (2}) shows the function $r(t)$, which decreases
monotonically. The function $r(t) = {\lambda_{phys} \over \lambda_H}$
[given by eq. (\ref{r})], describes the evolution of the physical wavelength
$\lambda_{phys}$ relative to the horizon wavelength $\lambda_H=1/h$.
Note that $r(t)$ remains below the unity
for $x > 0.001$. Hence, relativistic causality only should be violated on
transplanckian temporal scales (i.e., for $x < 0.001$). The origin of this
partial violation should be of quantum mechanical nature.

\subsection{Estimation of $\left<\varphi^2_L\right>$}

The equation (\ref{mot}) is very difficult to solve for a time
dependent power-law as (\ref{pow}). Hence, to make an
estimation of $\left< \varphi^2_L\right>$ for different stages of the
evolution of the universe we shall consider that $p$ is nearly
constant (i.e., ${\dot p\over p} \ll p/t$).
By solving the eq. (\ref{mot}), we obtain
\begin{equation}
\bar\xi_{k_R}[y(t)] = i \sqrt{\frac{\pi}{4(p-1)}} \sqrt{\frac{t}{t_0}}
{\cal H}^{(2)}_{\nu}[y(t)],
\end{equation}
where $y(t) = {k_r t^p_0 t^{1-p}\over p-1}$, $\nu = {p \over 2(p-1)}$,
${\cal H}^{(2)}_{\nu}$ is the second kind Hankel function
and $t_0$ corresponds to time when inflation begins. In order to
obtain $\left<\varphi^2_L\right>$ from eq. (\ref{gens}), it is necessary to
find $\bar D_{11}$, which is given by
\begin{equation}
\bar D_{11}(t) = \frac{\epsilon^3 \dot F k^2_0}{4\pi^2} \left|
\bar\xi_{\epsilon F}\right|^2 \simeq \frac{\epsilon^2}{16\pi^2}
\frac{(p-1)(p^2-1)}{t^{3p+1}_0} \  t^{2p-1},
\end{equation}
where we have made use of the fact that
$k^2_0(t) = {t^{2(p-1)}\over 4t^{2p}_0}
\left(p^2-1\right)$ (which is negative for $p<1$),
and the small argument asymptotic expansion for the
second kind Hankel function.
Under this approximation, the expectation value for
the second momenta of $\varphi_L$ is
\begin{equation}
\left<\varphi^2_L\right>
\simeq \left[ \frac{\epsilon^2}{16\pi^2} \frac{(p-1)(p^2-1)}{
2p t^{3p+1}_0} t^{2p} + C\right] \left(\frac{t}{t_0}\right)^{-(3p+1)},
\end{equation}
where $C$ is an arbitrary constant of integration. Notice that for $p=3/2$ it has
a singularity. One interesting case is $p=1/2$, that corresponds
to the radiation dominated universe. In this case $\left<\varphi^2_L\right>$
is given by
\begin{equation}
\left<\varphi^2_L\right> \simeq \left[\frac{3\epsilon^2}{2^7 t^{5/2}_0}t
+C\right] \left(\frac{t}{t_0}\right)^{-5/2}.
\end{equation}
Notice that $\left<\varphi^2_L\right>$ decreases with the time for
all values of $p$.
It is important to note that this approach is only valid when the
condition (\ref{ju}) is fulfilled.
Note that $k^2_0 >0$ for $p>1$ and $k^2_0<0$ for $p<1$. However
$\dot F>0$ for $p>1$ and $\dot F <0$ for $p<1$, so that
$\bar D_{11} > 0$, for all values of $p$. This means that
the number of degrees of freedom for $\chi_L$ is always increasing.
However the relevant field for us
is $\varphi_L$. Note that its number of degrees of freedom is
always increasing but its rate of increment is now
decreasing because of
the present day (and future) acceleration of the universe.
Hence, the model predicts a decreasing
rate in the increment of the
number of degrees of freedom, which (for $t\rightarrow \infty$)
will be almost constant.
Taking $p$ almost constant from
(\ref{ju}) we obtain that
\begin{equation}
\frac{1}{4} \ll (p-2)^2.
\end{equation}
This means that the stochastic approach here developed is
very efficient for $p \gg 2$ and has a reasonably good
behavior for $p<1$ (we are dealing only with $p>0$).
However, for $p \simeq 2$, the stochastic noise $\gamma $ in eq. (\ref{58})
should be taken into account.

\subsection{Calculation of the background field $\phi_c(t)$}

In order to estimate the evolution of the squared inflaton fluctuations
in different epochs of the evolution of the universe, we can make
a semiclassical approach $\varphi(t,\vec R) = \phi_c(t) +\phi(t,\vec R)$,
such that $\phi_c(t) = \left<\varphi\right>$ and $\left<\phi\right>=0$.
Hence,
\begin{equation}
\left<\varphi^2\right> = \phi^2_c(t) + \left<\phi^2\right>.
\end{equation}
To estimate $\left<\phi^2\right>$, we need to know $\phi_c(t)$, which
is the zero mode solution of the differential equation (\ref{mo}) and
has the general solution
\begin{equation}
\phi_c(t) = e^{-\int p(t)/t dt}
\left(A \  e^{-\int p(t)/t dt} + B\right) ,
\end{equation}
where $A$ and $B$ are constants of integration.
For the time dependent power-law (\ref{pow}), the solution
is
\begin{equation}
\phi_c(t) = t^{-\left(\frac{b^2}{4a}+\frac{2}{3}\right)}
e^{1.8 \left(\frac{a}{n}\right)t^{-n}-3.6 \left(\frac{b}{n}\right)t^{-n/2}-ct}
\left[A t^{-\left(\frac{b^2}{4a}+\frac{2}{3}\right)}
e^{1.8 \left(\frac{a}{n}\right)t^{-n}-3.6 \left(\frac{b}{n}\right)t^{-n/2}-ct}
+ B\right].
\end{equation}
Note that $\phi_c(t)$
decreases monotonically during all the history of the universe.

\subsection{Estimation of squared inflaton fluctuations
on the infrared sector}

Once known $\phi_c$, which becomes negligible for
late times, we can estimate the squared inflaton fluctuations
$\left<\phi^2_L\right>$ on the infrared sector. For late times one
obtains
$\left<\varphi^2_L\right>_{ t\gg G^{1/2}} \simeq
\left<\phi^2_L\right>_{ t\gg G^{1/2}}$, where
\begin{equation}
\left<\phi^2_L\right>_{t \gg G^{1/2}}
\simeq \left[ \frac{\epsilon^2}{16\pi^2} \frac{(p-1)(p^2-1)}{
2p t^{3p+1}_0} t^{2p} + C\right] \left(\frac{t}{t_0}\right)^{-(3p+1)}.
\end{equation}
Notice that the expectation value for the inflaton field fluctuations
decreases in all the epochs of the expansion of the universe,
independently of the $C$-value.

\section{Final comments}

In this paper we have studied a nonperturbative scalar field governed
cosmological model from a noncompact Kaluza-Klein theory of gravity
from a 5D apparent vacuum. This vacuum is defined as a purely kinetic
density Lagrangian for a scalar field minimally coupled to gravity
on a 5D background globally flat ($R^A_{BCD}=0$) metric.
We have worked a cosmological model with a time dependent power-law
$p(t)$ which describes an universe that initially is accelerated and
suffers an inflationary expansion ($p>>1$). However, this power
decreases with time through its minimal value $p \simeq 1/2$, which
describes a radiation dominated universe. Thereafter, this power begins
to increase passing by a matter dominated ($p \simeq 2/3$) epoch, to
the present day quintessential expansion with $p \simeq 1.2$.
The model predicts that $p$ will continue increasing and finally
will expand with $p \simeq c t$. This means that finally the universe
will adopt a de Sitter expansion. At this moment, the field $\phi_c(t)$
will tend to zero, when the system will adopt its minimum energetic
configuration.

We agree with the suggestion that perhaps some scalar (inflaton
field) has been sliding down its potential energy hill on a time scale
of billions of years rather than fractions of second (i.e., only during
inflation)\cite{kichen,EPJ04,gi27,guth}.
We obtain that the mass of the inflaton field is of the order of the Hubble
parameter and the scalar potential is quadratic in $\varphi$. Notice
that this potential [see eq. (\ref{pot2})] is induced geometrically and
take different representations in different frames. In our case, we are
dealing with a frame which is comoving with the expansion of the
universe.
In agreement with the
power-law (\ref{pow}), in the future the universe will be again dominated by
vacuum ${\rm p} \simeq - \rho$, to die in a de Sitter (inflationary)
expansion.
The stochastic approach here developed has a good behavior for $p \gg 2$
and $p<1$, but not for $p\simeq 2$. In that case, a more
complete stochastic formalism should be developed, in order to include
the stochastic noise $\gamma $ in the evolution of $\left<\varphi^2_L\right>$.

Finally, in our cosmlogical model the only free parameter is a
cosmological observable; the Hubble parameter $h$. In this work we are
dealing with some example given by eq. (\ref{pow}), but in general,
the theory can make predictions from observation for the
reconstruction of the Hubble parameter, which is expected from the
future SNAP data\cite{sta}.

\vskip .3cm
\noindent
\centerline{{\bf Acknowledgements}}
\vskip .1cm
\noindent
MA and MB acknowledge CONICET (Argentina) and UNMdP for financial
support.
JEMA acknowledges CONACyT (M\'exico) and IFM for financial
support.

\vskip 7cm
\noindent
{\rm Fig.1:}  Evolution of $p[x(t)]$
as a function of $x(t) = {\rm log}_{10}(t)$.\\
\vskip 7cm
{\rm Fig.2:}  Evolution of $r[x(t)]$
as a function of $x(t) = {\rm log}_{10}(t)$, during
the early stages of inflation.\\

\end{document}